\documentclass[sigconf, screen]{acmart}

\usepackage{wrapfig}

\AtBeginDocument{%
  \providecommand\BibTeX{{%
    \normalfont B\kern-0.5em{\scshape i\kern-0.25em b}\kern-0.8em\TeX}}}

\usepackage{xcolor}
\usepackage{subcaption}

\newcommand{\tool}{\textit{inSearch}}

\copyrightyear{2022} 
\acmYear{2022} 
\setcopyright{acmcopyright}\acmConference[CHI '22 Extended Abstracts]{CHI Conference on Human Factors in Computing Systems Extended Abstracts}{April 29-May 5, 2022}{New Orleans, LA, USA}
\acmBooktitle{CHI Conference on Human Factors in Computing Systems Extended Abstracts (CHI '22 Extended Abstracts), April 29-May 5, 2022, New Orleans, LA, USA}
\acmPrice{15.00}
\acmDOI{10.1145/3491101.3519725}
\acmISBN{978-1-4503-9156-6/22/04}

\begin{document}




\title[\tool{}]{Who Will Support My Project? Interactive Search of Potential Crowdfunding Investors Through \tool{}}



\author{Songheng Zhang}
\affiliation{%
  \institution{Singapore Management University}
  \country{Singapore}
}
\email{shzhang.2021@phdcs.smu.edu.sg}

\author{Yong Wang}
\affiliation{%
  \institution{Singapore Management University}
  \country{Singapore}
}
\email{yongwang@smu.edu.sg}

\author{Haotian Li}
\affiliation{%
  \institution{The Hong Kong University of Science and Technology}
  \city{Hong Kong SAR}
  \country{China}
}
\affiliation{%
  \institution{Singapore Management University}
  \country{Singapore}
}
\email{haotian.li@connect.ust.hk}

\author{Wanyu Zhang}
\affiliation{%
  \institution{ByteDance Ltd}
  \city{Beijing}
  \country{China}
}
\email{zhangwanyu.wanwoo@bytedance.com}

\renewcommand{\shortauthors}{Songheng Zhang, Yong Wang, Haotian Li, and Wanyu Zhang}


\begin{abstract}



Crowdfunding provides project founders with a convenient way to reach online investors. However, it is challenging for founders to find the most potential investors and successfully raise money for their projects on crowdfunding platforms. A few machine learning based methods have been proposed to recommend investors' interest in a specific crowdfunding project, but they fail to provide project founders with explanations in detail for these recommendations, thereby leading to an erosion of trust in predicted investors. To help crowdfunding founders find truly interested investors, we conducted semi-structured interviews with four crowdfunding experts and presents \tool{}, a visual analytic system. \tool{} allows founders to search for investors interactively on crowdfunding platforms. It supports an effective overview of potential investors by leveraging a Graph Neural Network to model investor preferences. Besides, it enables interactive exploration and comparison of the temporal evolution of different investors’ investment details.

\end{abstract}

\keywords{Visual Analytics, Crowdfunding, Comparative Analysis}


\begin{teaserfigure}
\vspace{-1em}
\centering
  \includegraphics[width=0.94\textwidth]{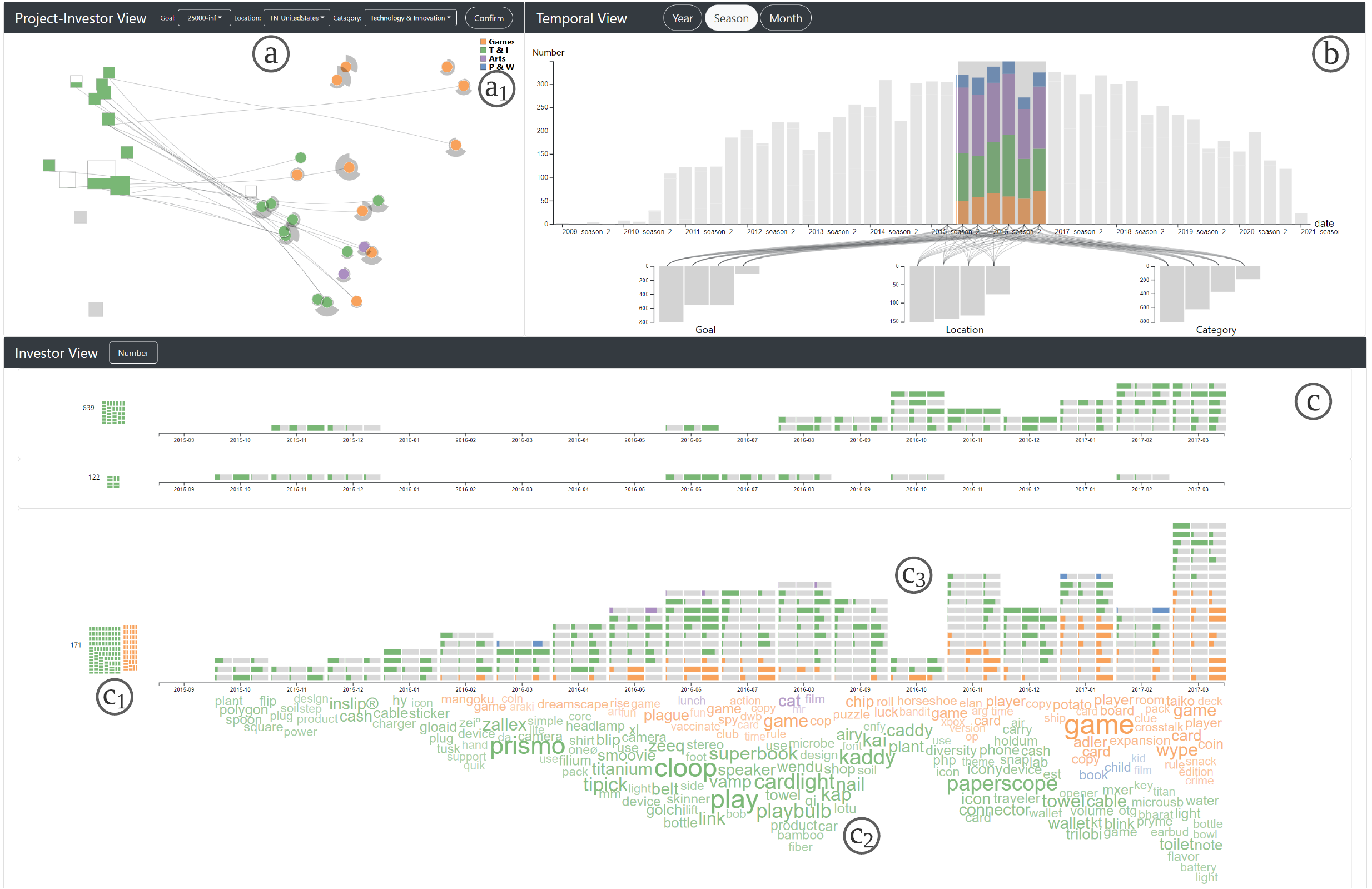}
  \vspace*{-5mm}
  \caption{\tool{} assists project founders in finding prospective investors for their projects: (a) the Project-Investor View allows users to manually select similar projects and automatically link to recommended investors; (b) the Temporal View displays investors' all investment projects and enables users to select a period of time on the timeline to analyze the invested project in details; three histograms summarize the distribution of the features of the invested projects within the selected time period; (c) the Investor View displays every recommended investors' projects which occurred at selected time;  (\textit{$c_1$}) the Treemap presents a characteristic proportional distribution of investments over selected time periods;  (\textit{$c_2$}) the Keyword Stream shows the keywords of invested projects; (\textit{$c_3$}) the Horizontal Bar delineates an invested project essential three features.}
  \label{fig:teaser}
\end{teaserfigure}

\maketitle


\section{Introduction}
Many crowdfunding platforms (e.g., Kickstarter and Indiegogo) have been launched in the past decade.
These platforms have become one of the most important ways for entrepreneurs and business owners to raise money for their business online, especially for startups and small businesses~\cite{yallapragada2011small,cressy2002introduction}.
Instead of using traditional methods of fundraising, crowdfunding relies on online platforms to access funds directly and quickly from Internet users by leveraging the power of the crowd~\cite{malone2009harnessing}.
Due to its significant convenience, crowdfunding has grown rapidly over the past decade. For example, the global crowdfunding market is valued at 12.27 billion U.S. dollars in 2021~\cite{MarketSize2020}.
Also, the global pandemic of COVID-19 has greatly affected many economic activities, especially small businesses, thereby increasing the demand for crowdfunding~\cite{covid-fueled-crowdfunding,macht2014benefits}.
Due to these reasons,
the crowdfunding market is expected to grow at a compound annual rate of over 16\% between 2021 and 2026~\cite{crowdfunding-market-21-26}.


However, not all crowdfunding projects are successfully funded:
$43\%$ of crowdfunding campaigns failed to receive a single investment and over $90\%$ of the campaigns did not achieve their fundraising goals~\cite{crowdfundingsuccessrate}.
The most common cause of fundraising failure is that founders fail to find and attract enough investors who are interested in their projects~\cite{an2014recommending,hui2014understanding}.
Therefore, it has become critical for project founders to identify the most potential investors.

Finding appropriate investors for a crowdfunding project is intrinsically a challenging task.
First, for different crowdfunding investors, \textit{\textbf{diverse factors}} can influence their decisions on whether backing a project or not, such as  project type, project updating frequency, fundraising goal, location of project founder, and other personal preferences of investors.
As there are so many factors affecting a project investment, it is difficult for a founder to take these factors into account and identify prospective investors.
Second, the preferences of investors can be \textit{\textbf{dynamically changing}}. Even for the same investor, the
risk tolerance, preferred project types, and other considerations may vary over time~\cite{guiso2018time}, making it challenging to find the most appropriate investors.
Third, project founders often have to identify potential investors
from social media platforms (e.g., Twitter and Facebook) or existing crowdfunding platforms. But there are often \textit{\textbf{a huge number of investors}} on those platforms, making it even difficult to recognize the most potential investors.

Our survey shows that a few studies have attempted to help project founders find potential investors. 
Specifically, these studies have applied machine learning techniques to recommend potential investors for a specific project.
For instance,
An~\emph{et al.}~\cite{an2014recommending} considered potential investors' social media activities and project-based features and further leveraged Support Vector Machine (SVM) to recommend investors for crowdfunding projects.
These studies partially addressed the first and third challenges above in identifying potential investors for a particular project.
However, they cannot explain why the recommended investors will be interested in the project and do not take into account the temporal changes of investors' investment preferences.


In this study, we propose \tool{}, an~\underline{in}teractive visual analytic system to support \underline{in}vestor \underline{Search}.
We work closely with four crowdfunding experts to collect their opinions about investors' search and iteratively refine our visual designs.
To model investors' heterogeneous investment characteristics, we present a novel graph neural network (GNN) based technique to incorporate multiple factors that can indicate investors' investment preferences.
Additionally, \tool{} allows project founders to compare and select desirable investors with three levels of investor details.
The Project-Investor View (Figure \ref{fig:teaser}a) provides an overview of the projects and corresponding investors and enables founders to discover the similarity of those items based on their distance, where clustered items indicate similar projects or investors.
The Temporal View (Figure \ref{fig:teaser}b) displays the temporal evolution of projects invested by investors.
The Investor View (Figure \ref{fig:teaser}c) shows the temporal details of projects invested by an individual investor, where a visualization design combining word cloud and stream graph is proposed to visualize invested projects' details.
%
%
%
%
With \tool{}, project founders can find the potential investors interested in their projects quickly and with more confidence.
In this paper, we evaluate \tool{} on the real-world datasets collected from Kickstarter, one of the most popular crowdfunding platforms, and display a preliminary usage scenario to demonstrate the usefulness and effectiveness of our approach.



The major contributions of our approach can be summarized as follows:
    
\begin{itemize}
    \item We present \tool{}, an interactive visual analytic system to assist project founders in achieving effective and confident search of potential investors on crowdfunding platforms, where a new and effective temporal visualization is designed to display the temporal evolution of investment preference of an investor.
    
    \item We show one usage scenario on the Kickstarter dataset to demonstrate the usefulness and effectiveness of \tool{}. 
    
    

\end{itemize}

\section{Crowdfunding Fundraising Analysis}
\label{sec:background}
Before designing \tool{}, we conducted semi-structural interviews with four crowdfunding experts
to understand the fundraising workflow and challenges they face. We contacted experts (\textit{E$_1$}, \textit{E$_2$}) via emails. We conducted remote interviews with experts (\textit{E$_3$}, \textit{E$_4$}) via Zoom. We asked all the experts the same questions.
These questions mainly focus on three aspects:
1) the general workflow of starting a crowdfunding project; 2) their suggestions for a successful crowdfunding project; 3) the challenges they have faced when launching a crowdfunding project.

\textbf{Participants.}
All participants launched at least two crowdfunding projects. One expert (\textit{E$_1$}) is the CEO of a crowdfunding agency that has helped project founders raise tens of millions of dollars. An expert (\textit{E$_2$}) has authored a handbook on crowdfunding fundraising.
He interviewed other crowdfunding founders who raised money ranging from thousands of dollars to millions and has participated in several crowdfunding projects. Expert (\textit{E$_3$}) launched three crowdfunding projects about technology and two of them succeeded. Expert (\textit{E$_4$}) had two crowdfunding projects about mother \& baby care. 


The expert feedback
reveals one key factor for a project to succeed as well as three design requirements associated with it. 
According to the feedback of experts, the most challenging and significant factor for successful crowdfunding is to find enough investors interested in the crowdfunding project once the project is launched. 
They spend much time determining potential target investors and further contacting them for fundraising. Such a key factor has led to three specific requirements:

\textbf{R1. Narrow down prospective investors who are interested in the project.} There are a lot of investors on the Internet, but it is impossible to contact
such a large number of investors individually. Hence, the system should recommend a set of prospective investors according to founders' project characteristics, filtering out investors who are unlikely to be interested in the current project.

\textbf{R2. Get a quick overview of a group of similar investors' general preferences.} Because there are numerous investors on the Internet, it is impossible to conduct a detailed analysis of each investor and then determine their willingness to invest. Therefore, it is important to help founders determine which investors are likely to invest in their projects and to analyze these investors thoroughly. If a group of investors is not interested in founders' projects, founders can stop studying them and turn to another group of investors. For example, from the timeline, if these investors have not invested in any projects that are similar to the founders' projects, it is clear that these investors are not interested in their projects.

\textbf{R3. Explore and compare the detailed investments of individual investors.} According to Behavior Finance, different investors' investment styles are not identical~\cite{shiller2003efficient,Kosiur01}. For example, some investors change project categories frequently and mainly focus on small-size projects. On the other hand, some investors are not active in the crowdfunding market, but they focus on one specific category of fund projects with high goals. Founders hope to intuitively explore and compare the detailed investments of individual investors, and then determine whether those investors match their projects.
An investor may invest in different projects over time. Hence, it is crucial to visualize the temporal evolution of investments by different investors.

\section{System Overview}


We propose \tool{}, an interactive visual analytics tool to help project founders identify investors' preferences. The system consists of two main modules: Graph Neural Network (GNN) model similarities between investors; interactive visualization which enables founders to determine selected investors' preferences.

\textbf{Dataset.} Our dataset is collected from Kickstarter, the popular crowdfunding platform. It contains the information of projects and their investors. The project features include projects' \textit{Fund Goal}, \textit{Category}, \textit{Updated Information}, \textit{Project Founder's Location}, \textit{Textual Description}, \textit{Comments}, and \textit{Launching Time}. As for investor features, the database includes their \textit{Invested Projects} and \textit{Location}.

\begin{figure}
    \centering
    \includegraphics[width=0.6\linewidth]{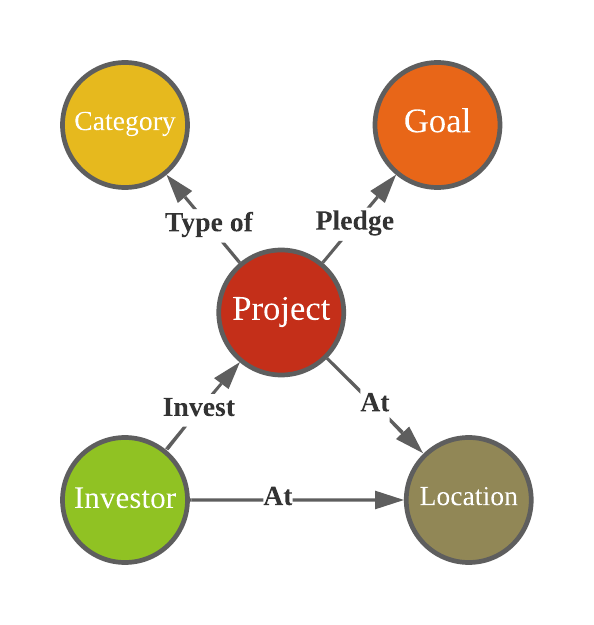}
    \caption{Projects, investors and their features are represented by graph vertices; their interactions are represented by edges. }
    \label{data graph}
\end{figure}

\textbf{Graph Neural Network.} Crowdfunding leverages the power of crowds.
Therefore, there are different communities of investors and these communities are crucial to the success of projects~\cite{tan2021crowdfunding}. 
Inspired by a previous study where GNN 
has been applied to detect communities in a Facebook social network~\cite{ugander2011anatomy}, we also apply GNN to identify investor communities in our crowdfunding dataset. 
To utilize GNN, we have constructed a graph as a crowdfunding graph. Unlike the Facebook social network, the crowdfunding graph does not explicitly label relationships between investors. But if they invested in the same projects, they will be considered as linked. 
In the crowdfunding graph shown in Figure \ref{data graph}, projects and investors are represented by the graph vertices. 
A graph vertex also encodes specific features of a project or investor, and these features are connected to the vertices of the project and investor.
Thus, by combining investor and project features with hidden communities, GNN can capture the investment preference similarity among different investors.

The crowdfunding graph is heterogeneous because investors, projects, and features differ~\cite{sun2012mining}. Meanwhile, RGCN, a GNN-based model, is well-suited for a heterogeneous graph with multiple relationships~\cite{schlichtkrull2018modeling,li2020peer}. Thus, we apply RGCN to the crowdfunding graph to represent investor preferences in terms of embedding, a high-dimensional vector. To find a group of investors with similar preferences, we use Euclidean distance to measure the investment preference similarity.

\textbf{Interactive Visualization.} Although machine learning based methods~\cite{an2014recommending,rakesh2016probabilistic,wang2020recommendation,wang2020bipartite} can efficiently predict potential investors for a specific project, they cannot explain why those investors are predicted. 
Contrarily, we provide founders with interactive visualizations that allow them to analyze potential investors intuitively based on the levels of detail they select.
For example, by inspecting the Project-Investor View, the founders can identify investors who have similar interests to projects. Then the Temporal View enables founders to track investors' interests over time. Founders use Investor View to determine which investors are most likely to invest in their projects.



\section{Visualization}
\tool{} is designed to help users intuitively explore and compare investor preferences at different levels of detail.
It consists of three views: Project-Investor View, Temporal View, and Investor View, displaying the investment histories of potential investors.

\subsection{Project-Investor View}
The Project-Investor View (Figure \ref{fig:teaser}a) shows the potential investors who may be interested in the founder's project. A founder can find existing projects that are similar to his/her own and then find investors who have invested in these similar projects. Project-Investor View displays the main characteristics of potential investors in terms of Investor Glyph. The relative distance between investors represents their preferences. After observing investors' characteristics and similarities, the founder can select a subset of investors for deeper analysis.


\textbf{Investor Glyph.} To facilitate better identification of  investors' characteristics, we design the investor glyph that encodes an investor's essential information about preferences. As shown in Figure \ref{fig:investor glyph}, the inner circle denotes the largest category of projects in an investor's investment. Therefore, users can straightforwardly recognize the project category an investor prefers. The outer circle in the glyph encodes the attributes of invested projects. And these encoded attributes indicate respectively the average goal amount of invested projects, the number of investment projects, and the average reward level of invested projects. The color of the outer circle is the same as that of the inner circle, but the saturation is different. In addition to preserving the main characteristics of an investor, the color consistency also helps the user to distinguish between the investor's category preference and the characteristics of invested projects. Therefore, the glyph enables the user to know the essential characteristics of investors.

\begin{figure}
        \centering
     \begin{subfigure}[b]{0.33\linewidth}
         \includegraphics[width=\linewidth]{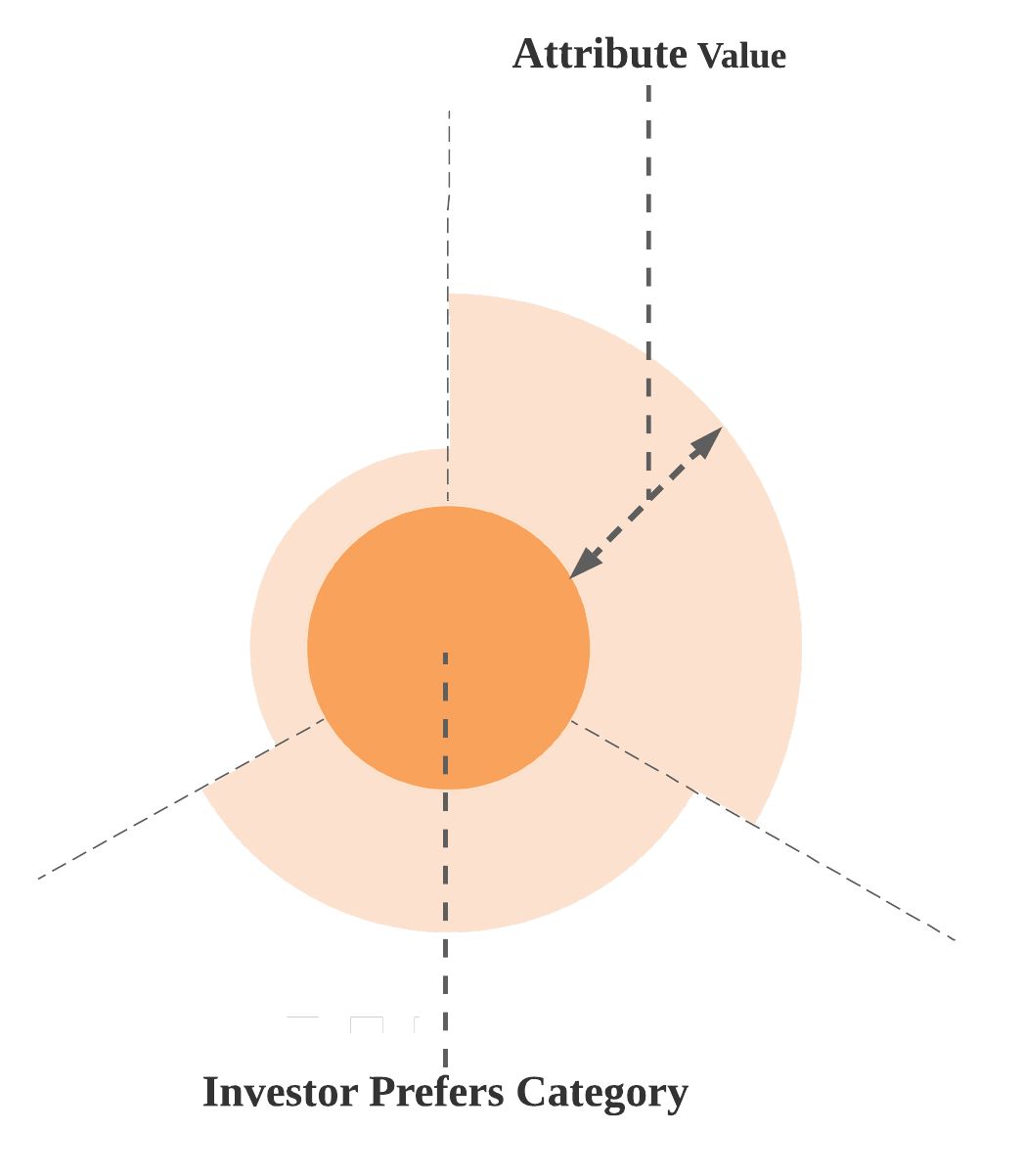}
         \caption{}
         \label{fig:investor_glyph}
     \end{subfigure}
     \begin{subfigure}[b]{0.33\linewidth}
         \includegraphics[width=\linewidth]{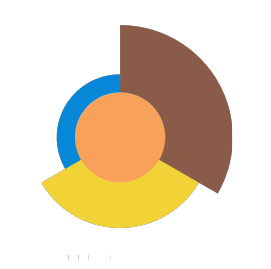}
         \caption{}
         \label{fig:alt_glyph 1}
     \end{subfigure}
     \begin{subfigure}[b]{0.3\linewidth}
         \includegraphics[width=\linewidth]{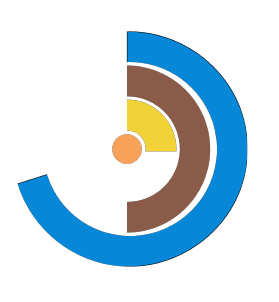}
         \caption{}
         \label{fig:alt_glyph_2}
     \end{subfigure}
        \caption{The investor glyph encodes an investor's three attributes: average goal amount, number of invested projects and average reward levels in clockwise order. (a) is applied in \tool{}; (b) and (c) are alternative glyph designs.}
\end{figure}

\textbf{Glyph Alternatives.} We have also explored other alternative designs when designing the investor glyph. In the first attempted glyph design, as shown in Figure \ref{fig:alt glyph 1}, each attribute is represented by a different color. We attempted to incorporate the design into our system. However, when similar investors are clustered together in the Project-Investor View, it is difficult for founders to identify each investor's characteristics since multiple colors are mixed in a limited space. The founder finds it difficult to process the cluttered visual information. In the second alternative glyph design shown in Figure \ref{fig:alt glyph 2}, the alternative design differs from our current glyph in its visual coding of invested project features. The alternative glyph uses inner circles' perimeter to represent the value of invested project features. Nevertheless, the difference in radius is difficult for founders to accurately perceive.


\textbf{Project-Investor View Layout.}
We apply UMAP~\cite{mcinnes2018umap} to downscale high-dimensional investors' embedding and project them onto the 2D space. The investors' relative similarity is preserved. The investors with similar preferences are close together on the Project-Investor View space, while those with large differences are far apart. Thus, founders can visually discover clusters and outliers (\textbf{R1}).

\subsection{Temporal View}

It is crucial to understand investors' long-term preferences while being able to recognize changes in their preferences in terms of investments. Hence, we design the Temporal View (Figure \ref{fig:teaser}b) which provides founders an overview of investors' historical investments. With this view, founders can detect changes in investors' preferences by observing the color changes of bars on the histogram. Also, founders can obtain investors' long-term preferences (\textbf{R2, R3}).

The Temporal View exhibits investors' complete investment history interactively. As shown in Figure \ref{fig:teaser}b, the timeline is placed along the x-axis and each scale tick represents a specific time interval for seasons. In Project-Investor View, founders have selected a set of investors and their investment projects will be arranged to different time intervals depending on launch time. We use a bar to encode these projects which are in a specific period. The height of each bar represents the number of projects within the time interval. The scale on the y-axis corresponds to the number of projects. A bar is composed of four small bars with different colors stacked vertically. These colors represent the project category in Figure \ref{fig:teaser}a$_1$. The height of each small bar also encodes the project number of one category. In addition, three histograms will appear when a founder does selection on the timeline histogram as Figure \ref{fig:teaser}b shows. Goal histograms delineate the distribution of project goal in terms of four levels (0-1,000, 1,000-10,000, 10,000-250,000 and 250,000-infinite). The Location histogram depicts the top five cities with the most launched projects. And the Category histogram represents the number of investment categories within selected time intervals.


\subsection{Investor View}

While the Project-Investor View and Temporal View provide founders with a quick overview of recommended investors, they only offer general information about investors as a group. For example, founders know the investor group preferences about project categories and goal amount. But founders also want to know the preferences of each investor. Thus, in the Investor View shown in Figure \ref{fig:teaser}c, the founder can analyze the characteristics of each project from each investor. By comparing the characteristics of past investment projects, the founders can determine whether an investor would be interested in their projects (\textbf{R2, R3}).

\begin{figure}
     \begin{subfigure}[b]{0.49\linewidth}
         \includegraphics[width=\textwidth]{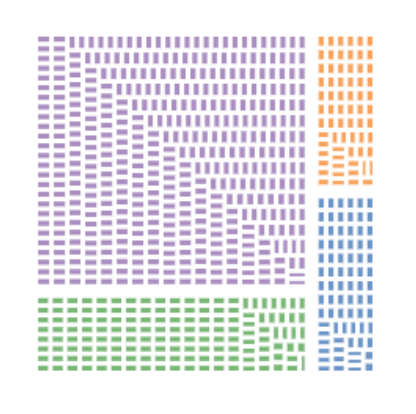}
             \caption{Category treemap}
         \label{fig:category treemap}
     \end{subfigure}
     \begin{subfigure}[b]{0.49\linewidth}
         \includegraphics[width=\textwidth]{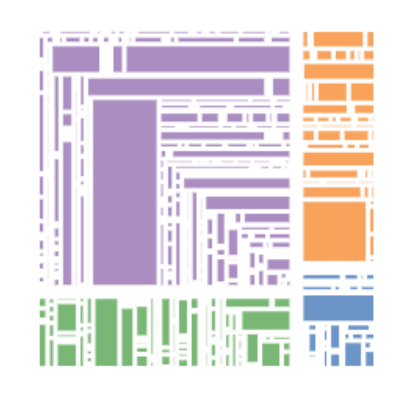}
         \caption{Goal treemap}
         \label{fig:goal treemap}
     \end{subfigure}
     \caption{An overview of an investor's investment project portfolio. Each small rectangle encodes a project. Rectangle color denotes project category. (a) Rectangles have the same size. (b) The size of each rectangle depends on its goal amount.}
     \label{fig:treemap}
\end{figure}

The Investor View shown in (Figure \ref{fig:teaser}c) mainly displays the invested projects' essential features including goals, comments, updates, and textual description  (\textbf{Section 3}). Each project is encoded by a project glyph that contains three small horizontal bars (Figure \ref{fig:teaser}c$_2$). Like the timeline histogram, the color of the project glyph represents the category of the project (Figure \ref{fig:teaser}a$_1$). Therefore, a founder can easily locate projects of interest. For example, a founder wants to discover some investment projects that belong to the same category as his/her project. Also, the length of three small horizontal bars delineates the value of the project attributes. Invested projects launched in the same time interval (e.g, 2016-10) are stacked vertically on top of each other over the timeline.

The projects' keywords are at the bottom of the timeline in Figure \ref{fig:teaser}c${_2}$. The size of keywords is consistent with the word frequency that happens in the project description. We use the method proposed in the Wordstream~\cite{dang2019wordstream} to prevent keywords from overlapping with each other. Besides, we use colors to encode the keywords. The color of the keywords represents categories of keywords' sourced projects. The keywords stream can assist a founder in finding projects of interest quickly. For example, a founder wants to raise money for a project about kitchenware. His project belongs to \textit{Technology \& Innovation}. He can check the keywords stream of investment projects and focus on the green font (green color encodes the category). If similar keywords appear (e.g., pan) the investor is likely to invest founder's project because the investor has invested in a similar one.

Treemap (Figure \ref{fig:treemap}) denotes a distribution regarding investments. The size of the treemap encodes the number of projects invested by investors. Depending on the category~(Figure \ref{fig:category treemap}) or goal amount~(Figure \ref{fig:goal treemap}), a treemap can display the distribution of projects. By clicking the treemap, the founder can change the encoded feature between category and goal. The color of each rectangle represents the corresponding category. In a category-based treemap, each rectangle size is equivalent. But in a goal-based treemap, the size of rectangles is dependent on projects’ fundraising goals.


\section{Finding Investors for a Small-size Project}

A usage scenario will illustrate the effectiveness and value of \tool{}.
Due to privacy reasons, the real names of investors are masked and replaced with numeric identifiers.

Suppose there is a project founder and he wants to find investors who will be interested in his kitchenware project. It is a small project with a low fundraising goal. Using the Project-Investor View, the founder can look for projects that are similar to his/her project and investors associated with these projects, after setting the project criteria. 
The founder recognizes that there are some groups of projects on the Project-Investor View. 
Because projects are grouped by their similarity in content and investors, the founder tries to select projects within a group. After selection, several lines automatically link selected projects with corresponding investors (Figure 1a). Additionally, the unselected projects turn gray to highlight what the founder is interested in. 

The Temporal View displays all invested projects from linked investors who appear on the Project-Investor View (Figure 1b). From the histogram above the horizontal timeline, the founder can get an overview of these investors' long-term preferences for projects. The founder wants to know the details of these invested projects that occurred in the last four seasons. Therefore, the founder selects the last four bins. Same as the Project-Investor View, unselected bins turn gray. With these histograms, the founder can get an overview of the features of these projects over the selected time period and find out whether they are similar to the founder's project. If not, the founder can select another cluster on the Project-Investor View.

After the founder finds a set of projects similar to his/her project in the Temporal View. The founder moves to the Investor View which helps the founder determine particular potential investors. And each row in the Investor View represents one investor. Using the treemap, the founder can first filter out some impossible investors (Figure 1c$_1$). In other words, the founder can focus on investors whose invested projects involve the category of the founder's project by treemap. After that, the founder looks at their previous invested projects. The founder clicks on one investor's treemap to see invested projects' keywords. By looking at the green font that corresponds to a kitchen product's category, the founder finds similar projects based on keywords (such as salt, cuisine, recipe, and sauce). After that, the founder identifies the possible investors in those projects and get their names on the left side of the row (for privacy reasons, we only display investors' IDs).

With the analysis from the overview to the details, the founder finally finds possible investors. The founder's project is more likely to be invested in if the investors have previously invested in similar projects since their previous investments demonstrate their interest in similar projects.

\section{Conclusion}

In this paper, we propose \tool{}, a visual analytics system to help project founders find potential crowdfunding investors interactively. 
We work closely with four experts
to summarize the major issues in finding interested investors, and then obtain the requirements for crowdfunding investor search.
Building upon this, we integrate RGCN, a GNN based model, with visualization techniques to provide project founders with the information of potential investors in multiple levels of detail.
Project founders can conveniently explore the overall features of potential investors and their past projects, and investigate the evolution and distribution of their investment interests.
We display one usage scenario to evaluate the \tool{} utility.

In future work, we would like to further evaluate our approach on the dataset collected from other crowdfunding platforms (e.g., Indiegogo) and conduct expert interviews.
Also, \tool{} currently focuses on exploring the text descriptions of each crowdfunding campaign.  
It would be interesting to further analyze the videos and images of crowdfunding projects,
which may also indicate the investment preferences of different investors. 


\begin{acks}
We would like to thank the domain experts for their insightful feedback and the anonymous reviewers for their valuable comments.
We also thank Professor Zhiling Guo for the constructive discussions on crowdfunding.
\end{acks}

\bibliographystyle{ACM-Reference-Format}
\bibliography{sample-base}

\appendix

\end{document}